\documentclass{aa}  
\usepackage[colorlinks=true,
    linkcolor=blue,
    filecolor=magenta,
    urlcolor=cyan,
    citecolor=blue]{hyperref}

\usepackage{graphicx}
\usepackage{txfonts}

\usepackage{natbib}
\bibpunct{(}{)}{;}{a}{}{,} 
\begin{document}

   \title{Are lithium-rich giants binaries?\\  A radial velocity variability analysis of 1,400 giants}
   \titlerunning{Are Li-rich giants binaries?: RV variability analysis}

   \author{Matias Castro-Tapia
          \inst{1}\fnmsep\inst{2},
          Claudia Aguilera-G\'omez\inst{1},
          \and        
          Julio Chanam\'e\inst{1}}

   \institute{Instituto de Astrofísica, Pontificia Universidad Católica de Chile, Av. Vicuña Mackenna 4860, 782-0436 Macul, Santiago, Chile\\
              \email{micastro2@uc.cl}
         \and
             Department of Physics and Trottier Space Institute, McGill University, Montreal, QC H3A 2T8, Canada}
            
    \authorrunning{Matias Castro-Tapia et al.}         

   \date{Received XXXX ; accepted YYYY}
 
  \abstract 
{The existence of low-mass giants with large amounts of lithium (Li) in their surfaces has challenged stellar evolution for decades. One of the possibilities usually discussed in the literature to explain these Li-rich giants involves the interaction with a close binary companion, a scenario that predicts that, when compared against their non-enriched counterparts, Li-rich giants should preferentially be found as part of binary systems.} 
{We aim to assemble the largest possible sample of low-mass giants with well-measured Li abundances, to determine with high statistical significance the close binary fractions of Li-rich and Li-normal giants, and thus test the binary interaction scenario for the emergence of Li-rich giants.}
{We developed a method that uses radial velocities (RVs) at three different epochs to quantify the degree of RV variability, which we used as a proxy for the presence of a close binary companion. The method was tested and calibrated against samples of known RV standard stars and known spectroscopic binaries. We then assembled a sample of 1418 giants with available RVs from RAVE, GALAH, and \textit{Gaia}, as well as stellar parameters and Li abundances from GALAH, to which we applied our variability classification. We could determine an evolutionary state for 1030 of these giants. We also compared the results of our RV variability analysis with binarity indicators from the \textit{Gaia} mission.}
{When applying our methodology to the control samples, we found that the accuracy of the classification is controlled by the precision of the RVs used in the analysis. For the set of RVs available for the giants, this accuracy is 80-85\%. Consistent with seismic studies, the resulting sample of giants contains a fraction of Li-rich objects in the red clump (RC) that is twice as large as that in the first ascent red giant branch (RGB). Among RC giants, the fractions of Li-rich objects with a high RV variability and with no RV variability are the same as those for Li-normal objects, but we find some evidence that these fractions may be different for giants in the first-ascent RGB.  Analysis of binary indicators in \textit{Gaia} DR3 shows a smaller fraction of binary giants than our criteria, but no relation can be seen between Li enrichment and binarity either.}
{Our RV variability analysis indicates that there is no preference for Li-rich giants in the RC to be part of binary systems, thus arguing against a binary interaction scenario for the genesis of the bulk of Li-rich giants at that evolutionary stage. On the other hand, Li-rich giants in the RGB appear to have a small but measurable preference for having close companions, something that deserves further scrutiny with more and better data. Additional measurements of the RVs of these giants at a higher RV precision would greatly help in confirming and more robustly quantifying these results.}

   \maketitle

\section{Introduction}

Low-mass stars deplete lithium (Li) during their lives, starting with pre-main sequence burning. In the main sequence, nonstandard mixing depletes the surface Li abundance even
further. The Li content is then diluted in the convective envelope once the star enters its post-main sequence evolution. Considering this history, supported by observations of Li abundances of stars of different masses, metallicities, and Galactic environments, we expect first-ascent low-mass red giant branch (RGB) stars to have little to no Li. Moreover, there is additional Li destruction as the giant ascends the RGB and crosses the luminosity function bump \citep[e.g.,][]{Shetrone2019}.  Often associated with thermohaline mixing \citep[but see][]{TayarJoyce2022}, this episode of extra-mixing decreases light element abundance in stars, not only affecting Li, but also the C, N, and carbon isotope ratio. Current spectroscopic surveys of stars, such as \textit{Gaia}-ESO \citep[e.g.,][]{Magrini2021}, the GALactic Archaeology with HERMES (GALAH) \citep[e.g.,][]{Deepak2020, Martell2021}, and the LAMOST (Large Sky Area Multi-Object fiber Spectroscopic Telescope) Experiment for Galactic Understanding and Exploration (LEGUE) \citep[e.g.,][]{Cai2023, Mallick2023}, have shown that most of the giants are consistent with this expectation, showing low values of Li or upper limits.

Nevertheless, there is a small percentage of giants that show a large Li abundance. The number of Li-rich giants, first discovered by \citet{WallersteinSneden1982}, has greatly increased due to surveys and dedicated high-resolution studies, showing that a small fraction of each sample is Li-rich, with the exact number depending on the selection criteria and the exact definition of an abnormally high Li abundance \citep[e.g.,][]{daSilva1995, Balachandran2000, Reddy2005, Gonzalez2009, Ruchti2011, Monaco2011, Lebzelter2012, Kirby2012, MartellShetrone2013, DelgadoMena2016, Deepak2020, Martell2021}. 
A limit of A(Li)>1.5 has historically been used to identify Li-enriched giants, given the amount of depletion expected during the lower RGB, combined with the expected initial Li abundance of stars similar to the Sun. This classification scheme, however, has misinformed the study of the phenomenon of Li-rich giants for a long time: because of different initial conditions regarding Li content, as well as different (standard, i.e., the first dredge-up) depletion efficiencies during the RGB, giants with a different stellar mass should not be expected to display the same levels of Li content while on the RGB/RC, and therefore what can be considered Li-normal or Li-rich depends on stellar mass and also -- though less strongly -- on metallicity \citep{AG2016, Sun2022, Chaname2022, Tayar2023}.

The source of Li in these anomalous giants remains a puzzle for stellar evolution. Diverse mechanisms have been invoked to explain the high Li abundance of giants. Some of these require an external source for the Li, such as the engulfment of planets \citep{SiessLivio1999} or mass transfer from an asymptotic giant branch star that can produce Li by hot bottom burning \citep{SackmannBoothroyd1992}. Some other hypotheses consider Li production in the interior of the star through the Cameron-Fowler mechanism \citep{CameronFowler1971}, requiring the rapid transport of elements between the region of nuclear burning, where fresh $^{7}$Be or Li are produced, and the surface of the star. Depending on the efficient extra-mixing mechanism considered, and when this is triggered, it could explain Li-rich giants at different evolutionary phases, such as the upper RGB or RC \citep[e.g.,][]{Schwab2020, Denissenkov2023}. Additionally, some processes require the presence of an external source to trigger the internal production of Li, either the tidal interaction of a binary companion \citep{Casey2019} or the merger with a helium-core white dwarf \citep{Zhang2020}. The Cameron-Fowler scenario for Li enhancement needs a fast and efficient mixing mechanism to increase the Li abundance of the giant. \citet{Casey2019} suggest that differential rotation that has been enhanced by the presence of a binary companion could produce the efficient levels of internal transport needed for Li enrichment. The additional angular momentum and tidal interaction produced by this companion would not only increase the surface Li but also the surface rotation rate. An interesting prediction of this model is that most of the Li-rich giant stars that are in the core-helium burning stage should have a binary companion since the sudden contraction of the giant at the beginning of the helium burning further enhances the transport of angular momentum in the stellar interior. It would therefore be very interesting to determine the incidence of binaries among Li-rich giants.

The enhancement of Li in Li-rich giants has often been associated with other signatures in unusual giants, such as fast rotation \citep[e.g.,][]{Carlberg2012}, infrared (IR) excess \citep[e.g.,][]{delaReza2015}, and chromospheric activity \citep[e.g.,][]{Sneden2022}. So far, the correlation between high Li and IR excess seems to be weak \citep{Rebull2015}, even for clump giants \citep{Mallick2022}. While \citet{Martell2021} found evidence that there does seem to be some relation between rapid rotation and Li enrichment in giants, \citet{Tayar2023} find that such a relation must be weak. The ingestion of a substellar mass companion or the presence of a binary companion may increase the rotation rate of giants, as well as their Li content \citep{PriviteraA2016, PriviteraB2016, Casey2019}.

Given the above possibilities, plus the fact that an important fraction of stars are members of multiple systems, it becomes a pressing question to investigate the incidence of binaries among Li-rich giants. Binarity has been extensively studied as a mechanism that can trigger changes in surface abundances due to either direct mass transfer or tidal interaction processes, and not only in red giants \citep[e.g.,][]{Jorissen2019}. The rate of binarity among low-mass stars can vary from about $15$ to $50\%$ \citep[e.g.,][]{Raghavan2010, Dieterich2012, MoeDiStefano2017, Gao2017, Moe2019, Merle2020}. Some of these binaries, such as single-lined spectroscopic binaries, can only be identified due to periodic or large variations in their radial velocity (RV), and as such, they are usually confirmed to be close binary systems following their periodic change in RV over several years \citep[e.g., the SB9 catalog of][]{SB9}. These systems tend to show large RV amplitudes, varying from orders of $\sim1$ to $10^{3}$ km/s, and present periods of $\lesssim10^{4}$ days \citep[see Fig. 3 of][]{Pourbaix2004}. Thus, comparing RV from two or more different epochs has been used as an indicator of binarity \citep[e.g.,][]{Birko2019,Jofre2016,Jofre2023}. Since the current theories on Li-rich giants related to binary systems predict that they should have periods of $\lesssim 3000$ days \citep{Casey2019}, detecting RV variations can be applied to test such approaches.

Although binaries could be an important ingredient for understanding Li-enrichment in giants, there are only a few studies devoted to estimating the binary frequency of Li-rich giants. \citet{deMedeiros1996}, using 12 Li-rich giants from CORAVEL, report that they do not show a higher binary frequency. \citet{Ruchti2011} find no important variation for their eight Li-rich giants, although using only two RV epochs. \citet{Jorissen2020} monitored the RV of a sample of 11 Li-rich giants and another one of 13 normal, non-Li-rich giants that they used as a control sample to study correlations between Li enrichment, binary fraction, and  IR excess. They found a normal binary frequency among Li-rich giants and no correlation between the presence of dust and high Li. The only possible correlation they found was between Li and fast rotation solely in the most enriched giants. Given that they did not identify all Li-rich giants as binaries, the enhanced rotational mixing produced by a binary companion scenario, as an explanation for the origin of most Li-rich giants, is not supported by this work. Additionally, \citet{Goncalves2020} analyzed 18 stars, 17 of which are Li-rich giants, and found that some of them do have a RV variation, with five possibly having binary companions. For the remaining stars, there are only small changes in RV or not enough information to do a proper analysis. Given the lack of RVs for most of the stars, no other conclusions can be made regarding binarity and Li enrichment. Recently, \citet{Sayeed2024} found a similar rate ($\sim15\%$) of binary candidates for Li-rich and Li-normal giants when analyzing binarity indicators in \textit{Gaia}.

Since most of these results are still inconclusive, the binarity rate must be evaluated in larger samples of stars with a measured Li in different ranges of abundance to better constrain the proposed theories behind the Li enrichment in red giants. \textit{Gaia} Data Release 3 (DR3; \citealt{Gaia2023}) provides a new epoch of RV that can be used along with previous data to test binarity. Therefore, in this study, we combined information on red giants from \textit{Gaia} DR3, GALAH DR3, and the sixth data release (DR6) of the Radial Velocity Experiment (RAVE, \citealt{RAVEDR6}) to evaluate the existence of RV variability, as a proxy for the presence of a binary companion, and its relation with Li abundance.

This paper is organized as follows. In Sect. \ref{sec:2} we present the methodology we developed to determine whether a given target is highly likely to be on a close binary system or not based on RV variability and calibrate the method against well-known spectroscopic binaries and RV standards (stds). In Sect. \ref{sec:3} we explain how we applied our classification to all red giants that have RV measurements simultaneously in \textit{Gaia} DR3, GALAH DR3, and RAVE DR6, and for a sample of red giants with Li abundance reported in GALAH DR3. In Sect. \ref{sec:4} we compare the results obtained in Sect. \ref{sec:3} with information on binarity and variability in \textit{Gaia} DR3. Finally, our conclusions are summarized in Sect. \ref{sec:5}.

\begin{figure*}
        \includegraphics[width=18.0cm]{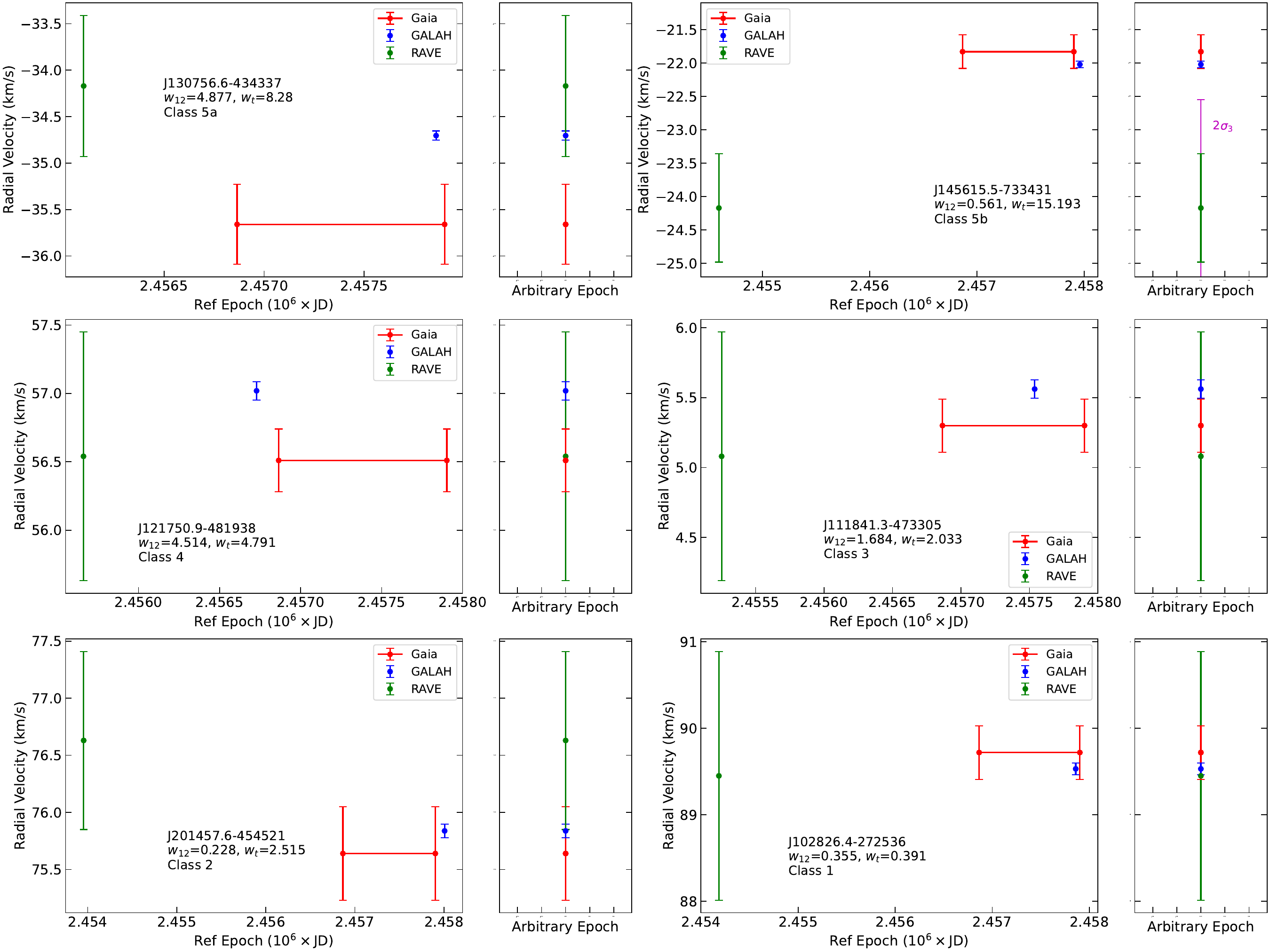}
    \caption{Each panel shows, for a star in each of the classes in Table \ref{tab:w12-wt}, the series of RV epochs used to determine the degree of RV variability of such a star (\textit{Gaia} DR3 in red, GALAH DR3 in blue, and RAVE DR6 in green). The left-hand panels show the time series including their RVs, reference epochs, target name, $w_{12}$ value, $w_{t}$ value, and the class assigned. For the \textit{Gaia} DR3 case, the value reported is the mean value of measurements from 25 July 2014 to 28 May 2017, which is the total period of time of the data processed for the data release 3 \citep{deBruijne2022}, and they are represented in the panels as a horizontal bar covering the time range. The right-hand panels show the three RV for an arbitrary epoch so that we can note how the error bars among the measurements are related for each class.}
   \label{fig:RV}
\end{figure*}

\section{Methods}\label{sec:2}
\begin{table*}
        \centering
        \caption{Definition of six classes of RV variability.}
        \label{tab:w12-wt}
        \begin{tabular}[H]{llll} 
                \hline
                Class & $w_{12}$ and $w_{t}$ values & Properties observed in the time series & Certainty of\\
                 &  & &variability\\
                \hline
                5a & $w_{12}\geq2$ and $w_{t}\geq5$ & The error bars of $RV_{1}$ and $RV_{2}$ clearly do not intersect with each other, & high\\
                & & and the error bar of $RV_{3}$ clearly does not intersect the error bars of &\\
                & &$RV_{1}$ or $RV_{2}$. &\\\\
                5b & $w_{12}<2$ and $w_{t}\geq11$ & The error bars of $RV_{1}$ and $RV_{2}$ intersect with each other, but the error & high\\
                & & bar of $RV_{3}$ clearly does not intersect the error bars of $RV_{1}$ and $RV_{2}$, and&\\
                & & they only could intersect if the error bar of $RV_{3}$ was $\gtrsim2\sigma_{3}$&\\\\
                4 & ($w_{12}\geq2$ and $w_{t}<5$) or & The error bars of $RV_{i}$ and $RV_{j}$ clearly do not intersect with each other & medium\\
                &($w_{12}<2$ and $3.3\leq w_{t}<11$)& but the error bars of pairs $RV_{i}$-$RV_{k}$ and $RV_{j}$-$RV_{k}$ intersect with each other;&\\
                & & considering $i,j,k=1,2,3$ with $i\neq j\neq k$.\\\\
                3 & $1\leq w_{12}<2$ and $w_{t}<3.3$ & The error bars of all $RV_{i}$-$RV_{j}$ pairs intersect with each other (for $i,j=1,$ & medium-low\\
                & &$2,3$ and $i\neq j$), but the pair $RV_{1}$-$RV_{2}$ only can fulfill $RV_{l}+\sigma_{l}<RV_{k}$ or &\\
                & & $RV_{l}-\sigma_{l}>RV_{k}$ (for $\sigma_{l}>\sigma_{k}$; $k,l=1,2$) &\\\\
                2 & $w_{12}<1$ ; $w_{t}-w_{12}\geq 1.5$&The error bars of all $RV_{i}$-$RV_{j}$ pairs intersect with each other (for $i,j=1,$ & low\\ &and $w_{t}<3.3$ &$2,3$ and $i\neq j$), the pair $RV_{1}$-$RV_{2}$ fulfill $RV_{l}-\sigma_{l}\lesssim RV_{k}\lesssim RV_{l}+\sigma_{l}$ (for $\sigma_{l}$ &\\
                & &$>\sigma_{k}$), also $\max{\{RV_{2}+\sigma_{2},RV_{1}}+\sigma_{1}\}>RV_{3}+\sigma_{3}$ or $\min{\{RV_{2}-\sigma_{2},RV_{1}}$ &\\
                & & $-\sigma_{1}\}<RV_{3}-\sigma_{3}$.&\\\\
                1 & $w_{12}<1$ and $w_{t}-w_{12}<1.5$ & The error bars of all $RV_{i}$-$RV_{j}$ pairs intersect with each other (for $i,j=1,$ & no variability\\
                & &$2,3$ and $i\neq j$), the pair $RV_{1}$-$RV_{2}$ fulfill $RV_{l}-\sigma_{l}\lesssim RV_{k}\lesssim RV_{l}+\sigma_{l}$ (for $\sigma_{l}$&\\
                & &$>\sigma_{k}$), also $\max{\{RV_{2}+\sigma_{2},RV_{1}}+\sigma_{1}\}<RV_{3}+\sigma_{3}$ and $\min{\{RV_{2}-\sigma_{2},RV_{1}}$&\\
                & &$-\sigma_{1}\}>RV_{3}-\sigma_{3}$.&\\\\
                \hline
        \end{tabular}
 \tablefoot{The classes were obtained by taking different ranges in the $w_{12}$-$w_{t}$ space parameter. The third column describes the features in the RV measurements $RV$ and errors $\sigma$ for each class from what was observed in the time series of a sample of heterogeneous red giants.}
\end{table*}

\subsection{RV variability as a binary indicator}\label{sec2.1}

One of the most used methods for detecting spectroscopic binary systems is the $F2$ statistic \citep[e.g.,][]{Jofre2016, Jofre2023, Jorissen2020}. For a star $j$ with $N_{j}$ radial velocities measurements $RV_{i,j}$, associated errors $\sigma_{i}$ for each epoch, and mean radial velocity $\overline{RV_{j}}$, a $\chi^{2}_{j}$ can defined as follows,
\begin{equation}\label{eq1}
  \chi^{2}_{j}=\sum^{N_{j}}_{i=1}\frac{(RV_{i,j}-\overline{RV_{j}})^{2}}{\sigma_{i}^{2}}. 
\end{equation}
The $F2$ technique is based on approximating the distribution of the $\chi^{2}$ values by a normal distribution of zero mean and standard deviation equal to 1. Since this method is used when having different amounts of observations per star, defining the $F2$ distribution allows the $\chi_{j}^{2}$  to be normalized by the degrees of freedom $\nu_{j}=N_{j}-1$. For each star, this is defined as follows \citep{WilsonHilferty},
\begin{equation}
  F2_{j}=\sqrt{\frac{9\nu_{j}}{2}}\left[\left(\frac{\chi^{2}_{j}}{\nu_{j}}\right)^{1/3}+\frac{2}{9\nu_{j}}-1\right].  
\end{equation}
Thus, significant RV variations correspond to those objects with high values of $F2$ that are far from the average 1 (usually $F2$>3 or >4). The reliability of this judgment requires having a large number of measurements per star $j$ to have a statistically significant mean $\overline{RV}_{j}$ in the definition shown in Eq. \ref{eq1}. Moreover, $\sigma_{i}$ is usually taken as a fixed stable value for all the RV measurements for a given star, which can be assumed as the long-term stability of the instrument. Hence, the $F2$ statistic, at least in its regular form, does not contemplate different RV errors with different instruments for the same star and would not be stable enough when using just a few epochs of RV from diverse surveys.

Another statistic employed to quantify RV variability as a proxy for identifying binaries involves computing the probability $P$ that two draws from different distributions have a large difference. In the RV variation context, $P$ can be understood as the probability that $RV_{2}$ with an error $\sigma_{2}$ is larger than $RV_{1}$ with an error $\sigma_{1}$ (where 1 and 2 are the indexes representing different instruments), and is usually expressed using the $\mathrm{erf}$ function as follows,
\begin{equation}
    P=\frac{1}{2}\left[1+\mathrm{erf}\left(\frac{RV_{2}-RV_{1}}{\sqrt{2(\sigma_{1}^{2}+\sigma_{2}^{2})}}\right)\right].
\end{equation}
Then, one can compute a statistic, $p_{log}$, defined as follows,
\begin{equation}
    p_{log}=-\log_{10}(1-P),
\end{equation}
 which increases when $P$ is closer to $1$. Thus, high variability is more probable for large values of $p_{log}$ \citep[see][]{Matijevic2011, Birko2019}. This technique is useful to compare only two RV measurements considering independent errors for each observation. However, while the probability rises for small values of $\sigma_{2}^2+\sigma_{1}^{2}$ (i.e., more precise RV measurements), an under or overestimation of one of these errors can lead to incorrectly classifying a star as RV variable or RV constant, respectively. 

In order to avoid this undesired behavior of the $p_{log}$ function, we defined a new statistic that allows us to generalize $F2$ and $p_{log}$ for the case of three velocity epochs and all from different instruments, since in this work, we use RVs from \textit{Gaia}, GALAH, and RAVE. First, we defined $w_{ij}$ as a variation parameter between two RV epochs,
\begin{equation}
    w_{ij}=\frac{(RV_{i}-RV_{j})^{2}}{(\sigma_{i}^{2}+\sigma_{j}^{2})}
,\end{equation}
where $i,j=1,2,3$ with $i\neq j$, and $RV_{i}$, $RV_{j}$, $\sigma_{i}$, and $\sigma_{j}$ are their respective RV measurements and errors. We also defined $w_{t}$ as the sum of the not repeated $w_{ij}$ elements (since $w_{ij}=w_{ji}$),
\begin{equation}
    w_{t}=w_{12}+w_{13}+w_{23}
.\end{equation}

\begin{figure*}
    \centering
    \includegraphics[scale=0.56]{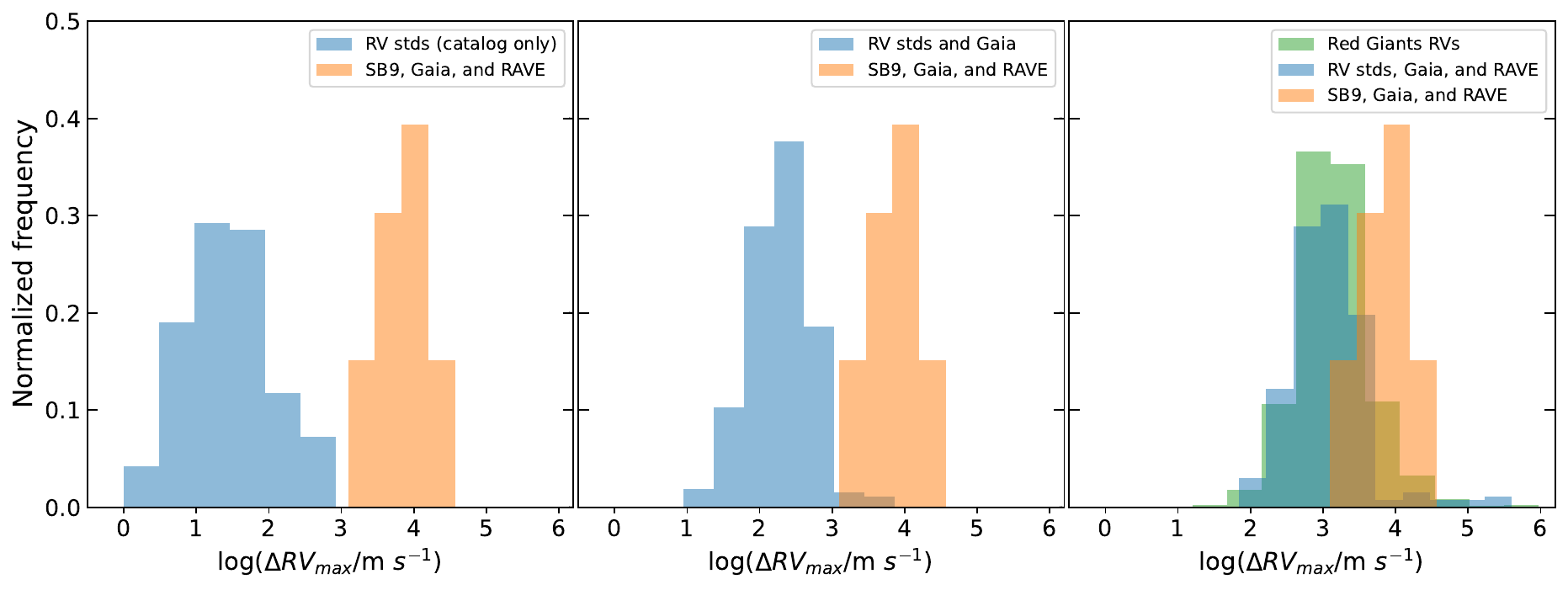} 
    \caption{Distributions of the maximum absolute difference in RV measurements ($\Delta{RV_{\mathrm{max}}}$) for the red giants analyzed in Sect. \ref{sec:3} (green), as well as the RV standard stars (blue) and SB9 binaries (orange) used to test the $w$ classification. For the red giants, the RV measurements are from \textit{Gaia}, GALAH, and RAVE. For the SB9 cases, the RVs are from the catalog itself, \textit{Gaia}, and RAVE. For the RV stds, each panel shows a different case: the first panel has RVs only from the catalog, the second panel includes the \textit{Gaia} RVs, and the third panel includes the \textit{Gaia} and RAVE RVs.}
    \label{fig:hist}
\end{figure*}
We choose to perform our variability analysis in a two-parameter space by considering $w_{12}$ and $w_{t}$, where $w_{12}$ was taken as the $w_{ij}$ combination that considers the two RV measurements with the smallest errors. Thus, $w_{12}$ allows less evident variations to be found, which can be confirmed by a high $w_{t}$. Secondly, if just a small change in RV occurs between the two best measurements, a third epoch that rises $w_{t}$ enough, can indicate a variation omitted by $w_{12}$, even when the third epoch has a larger RV error. Conversely, if there exists a variation according to $w_{12}$, but that is not high enough to confirm variability, $w_{t}$ can reject a star as a variable if its value is also low.

Now, we need to calibrate $w_{12}$ and $w_{t}$, that is, determine what values of these parameters correspond to a variable RV and what values to a constant RV. We started comparing the time series of a number of red giants with RV measurements reported in \textit{Gaia} DR3, GALAH DR3, and RAVE DR6 (see Sect. \ref{sec:3} for details on data selection). Once we identified a set of giants ($\sim 50$ targets) that constitute an apparently heterogeneous sample in terms of the differences among their RV measurements, we defined degrees of variation (classes) based on their plots of RV vs. time and their $w$ values. 
Of course, it is expected that there will exist a gray area of some size in between classes. In Fig. \ref{fig:RV} we show some examples of the run of RVs against time for a set of the stars used in this work, chosen to illustrate the full range of variability we see in our sample. For each representative example, we show the corresponding values for $w_{12}$ and $w_{t}$ and the variability classes assigned to each giant according to what is described in Table \ref{tab:w12-wt}.

To construct the variability classifications in Table \ref{tab:w12-wt}, we conducted a qualitative assessment of our set of giants and grouped them in different subsamples, considering the observable differences in the RV time series and the blending of error bars at arbitrary epochs (as shown in the right-hand subpanels of Fig. \ref{fig:RV}). By doing so, we defined distinct subsamples based on observable characteristics. We established representative thresholds in $w_{12}$ and $w_{t}$ that mostly enclose the features observed in each group of variability and we made sure that these thresholds described nonoverlapping zones in the $w_{12}$-$w_{t}$ while encompassing the entire parameter space.

Among the three surveys, RAVE usually has the largest RV errors, hence $w_{12}$ was selected as the $w$-parameter found comparing \textit{Gaia} and GALAH measurements, specifically, $RV_{1}=RV_{\mathrm{\textit{Gaia}}}$ and $RV_{2}=RV_{\mathrm{GALAH}}$.  We defined our 6 classes described in Table \ref{tab:w12-wt} based on different intervals in the $w_{12}$-$w_{t}$ parameter space, and a degree of certainty of variability was designated for each class as high, medium, medium-low, low, and no variability. We fixed the thresholds for each type of variability based on the separation of the RVs, and whether their error bars intersected with each other independently of the epoch. The high certainty of variability class was divided into 2 classes. For the case of a strong variation in the \textit{Gaia}-GALAH parameter, we defined the class 5a, whose interval takes values of $w_{12}\geq 2$ and $w_{t}\geq 5$ to fulfill a high certainty of variability. On the other hand, when $w_{12}<2$, we only established important variations and defined the class 5b for those stars with $w_{t}\geq 11$.

\subsection{Calibration against close binaries and RV stds}\label{sec2.2}
Our classes in Table \ref{tab:w12-wt} were created based on the observed characteristics of a set of RV time series chosen to be representative of the variety of cases present in the training sample used in Sect. \ref{sec2.1}. However, we do not know if the stars in this training sample are confirmed binaries or single stars.  Therefore, in this section we test our classification scheme against two sets of calibrating stars: 33 confirmed spectroscopic binaries from the SB9 catalog \citep[][]{SB9} and  $263$ stars from the RV standard catalog of \citet{Soubiran2018}. These stars all have multiple RV measurements in their respective catalog and have registered RV in \textit{Gaia} DR3 and RAVE DR6 as well. We chose to assign $RV_{1}$ and $RV_{3}$ to come from \textit{Gaia} and RAVE, respectively, and $RV_{2}$ is taken from the corresponding catalog. Thus, one star has more than one subset of RVs and $w$ values associated. For example, suppose that an SB9 star has $N_{r}$ RVs registered in that catalog, we can obtain $N_{r}$ subsets ($RV_{1}, RV_{2, i}, RV_{3}$), for $i=1,2...,N_{r}$ corresponding to each measurement in SB9, and recalling that $RV_{1}$ and $RV_{3}$ are fixed to the \textit{Gaia} and RAVE measurements. The total number of subsets obtained was $188$ for the SB9 catalog and $3477$ for the RV standard stars catalog.

The samples from the SB9 catalog and the RV stds are expected to be the cases where extreme RV variability and low to no variability are observed respectively. In using these samples for calibration purposes, we must consider that there is heterogeneity among them regarding the measurements. First, the average error in RV for the SB9 measurements is $\sim230$ m/s, while for the RV stds, it is $\sim2$ m/s. This supposes a difference of two orders of magnitude between the errors in the samples that would affect the values of $w_{12}$ and $w_{t}$, specifically, the small errors in the RV stds can increase both parameters leading to misclassification of these targets as variables. Moreover, as pointed out by \cite{Soubiran2018}, the stars in their catalog are considered RV stds above $300$ m/s. In light of these considerations, we fix $\sigma_{2}=300$ m/s for the RV stds, which would partially avoid assigning high variability to a target with RV variations under that level.

In Fig. \ref{fig:hist} we show the distribution of $\Delta RV_{\mathrm{max}}$, the maximum of the absolute RV difference found in the SB9 and RV stds samples, and also for the sample of red giants that we analyze in Sect. \ref{sec:3}. For the SB9 and the red giants, we only considered the RVs from their respective source catalogs, and the corresponding surveys (\textit{Gaia}-GALAH-RAVE measurements for the giants, see Sect. \ref{sec:3} for details). For the sample of RV stds, we show different combinations of the available RV measurements for that sample. The leftmost panel shows the distribution when only the RV measurements from the RV stds catalog are taken. In this first case, the distributions of the SB9 and the RV stds are well separated and do not overlap each other. This means that, with the RV precision of the RV stds source catalog, single stars can be well separated from close binaries like those in SB9 easily, just using $\Delta RV_{\mathrm{max}}$ and a few RV epochs. So if the giants that are the focus of this paper had RV measurements of that level of precision, the main question of this paper would be straightforward to solve with this method.

The middle and right-hand panels of Fig. \ref{fig:hist} show what happens when the \textit{Gaia} and RAVE RVs are included in the $\Delta RV_{\mathrm{max}}$ analysis of the sample of RV stds. As we add the \textit{Gaia} and RAVE data the $\Delta{RV_{\mathrm{max}}}$ of the RV stds move to larger values, and the distribution progressively overlaps with the SB9 region. This means that with the RVs available for the giants, we expect the classification with the $w$ values to have an uncertainty area where RV variable and non-variable objects will suffer some degree of misclassification.

\begin{figure}
    \centering
    \includegraphics[scale=0.5]{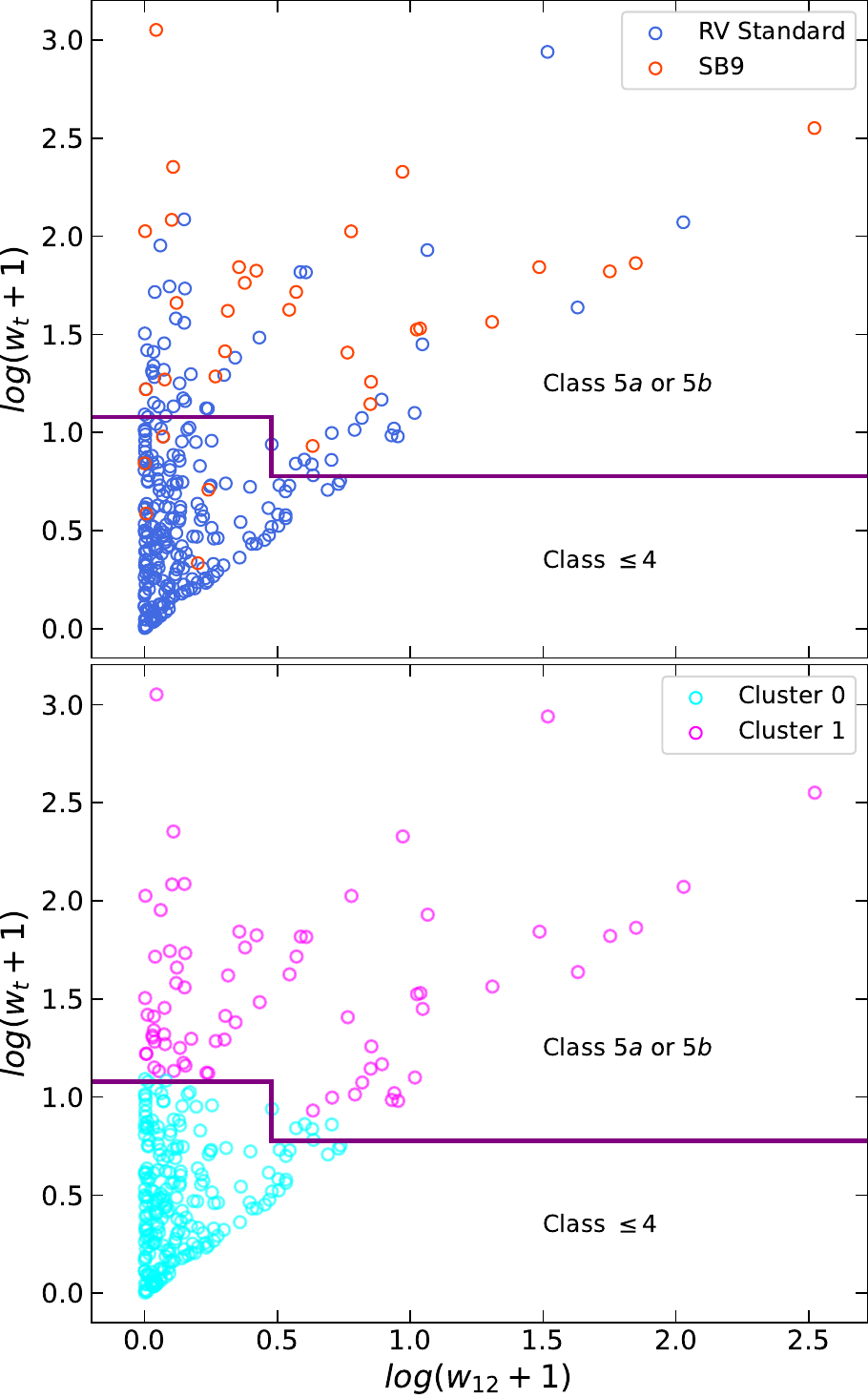} 
  \caption{The top panel shows the $w_{t}$-$w_{12}$ logarithmic distribution obtained for SB9 (orange) and RV-standard star (blue) subsets with \textit{Gaia} DR3, RAVE DR6 measurements, and one measurement obtained directly from SB9 or the RV standard stars catalog of \citet{Soubiran2018}. The bottom panel shows the $w_{t}$-$w_{12}$ logarithmic distribution for the same subsets shown in the top panel, but colored according to the cluster that the K-means algorithm decided they belonged to (i.e., cyan for stars in Cluster 0 and magenta for those in Cluster 1). For both panels, we added a purple line that indicates the separation between class $\leq 4$ and class 5a-b according to our classification based on the $w$ values.}
    \label{fig:stsb}
\end{figure}

We tested our classification on the control samples of RV stds and SB9 using the $w$ values for one subset per star, that is, only using one RV from each catalog, and then did a second test using all the subsets. The top panel of Fig. \ref{fig:stsb} shows how the subsets are distributed in the $w_{t}$-$w_{12}$ logarithmic space \footnote{We used $\log(w+1)$, thus the minimum value in the log space will be 0 as $\min(w)=0$ (for any $w$ defined).} considering just one (arbitrary) subset per star; the purple solid line indicates the separation between the star subsets considering class $\leq 4$ (below the line), and with high certainty of variability, that is, class 5a or 5b (above the line). We found that classification based on $w$ parameters indicates medium certainty of variability to no variability for about $80.2\pm5.5\%$ of the RV standard stars ($84.0\%$ when all standard-star subsets are used), while about $84.9_{-16.0}^{+15.1}\%$ of the SB9 stars have a high certainty of variability ($94.15\%$ when all SB9 subsets are used). 

This means that, for about $80$ to $95\%$ of the cases, the methodology is capable of distinguishing RV-variable and RV-non-variable stars down to a level of $300$ m/s. Since the SB9 spectroscopic binaries used in this calibration have orbital periods between $\sim$10-5000 days, it can be concluded that when we apply our method to the giants with Li measurements (Sect. \ref{sec:3}), we will be able to identify close companions to those giants with periods in that same regime with a success rate of $80$ to $95\%$. At the same time, there is a $20\%$ chance that the method flags as RV variable a star that is considered an RV standard down to 300 m/s, and between $5$-$15\%$ chance of misidentifying a spectroscopic binary as non-RV-variable. The former case might be produced by large RV errors masking a real RV variation facing our methodology, and the latter could be due to cases when the specific set of RV epochs entering the analysis are not sufficiently spread over the orbital period.  To account for this we would have to examine one by one all our targets, which we choose not to attempt because it beats the purpose of the statistic we construct, which can be systematically applied to a large collection of stars with RV measurements.

Next, we used a clustering algorithm to check whether the RV stds and the spectroscopic binaries are clearly separated into two groups in the $w_{12}$-$w_{t}$ parameter space. We used \texttt{KMeans} from \texttt{scikit.cluster} to apply the K-means clustering to the samples in the $w$-parameter space. Since the K-Means algorithm finds which point must belong to a certain group based on their Euclidean distance to the cluster centroids and the $w$ values in the samples can be $\gtrsim 10^{2}$, we considered finding the clusters in the logarithmic space of $w_{12}$-$w_{t}$, and the coordinates of a single point $i$ are $x_{i}=(\log(w_{12_{i}}+1),\log(w_{t_{i}}+1))$. We established that two clusters must be found, and the algorithm iterates defining two cluster centroids $\mu_{j}$ ($j=0,1$) until the cluster inertia $\phi$ has been minimized \citep{KMeans}, where $\phi$ is given by  
\begin{equation}
    \phi=\sum_{i=0}^{n}\min_{\mu_{j}}{(\|x_{i}-\mu_{j}\|^{2})}.
\end{equation}
Each point $x_{i}$ belongs to the cluster $j$ whose $\mu_{j}$ is the nearest in the minimization of $\phi$. The bottom panel of Fig. \ref{fig:stsb} shows the two clusters found in the logarithmic parameter space.

We find that $96.4\%$ of stars in Cluster 0 were class $\leq 4$ ($100\%$ for all subsets) according to our threshold for the $w$-parameter space, while $100\%$ of stars in Cluster 1 were class 5a or 5b ($93.7\%$ for all subsets). Therefore, the clustering algorithm naturally reproduces with high correspondence the separation between RV variable and non-variable stars achieved by the criteria in Table \ref{tab:w12-wt}. This validates the use of those criteria for our purposes of identifying binaries among giants with Li measurements. It is important to note that we consider three different classification groups: Those determined by the clustering algorithm (Cluster 0 or 1), those identified through extensive RV previous studies (SB9 or RV standards), and those classified based on the variability classes outlined in Sect. \ref{sec2.1}. While the clustering algorithm shows a separation similar to our threshold for high variability (class 5a-b vs class $\leq 4$), not all the objects categorized in Cluster 0 and 1 necessarily correspond to SB9 and RV standards, respectively. This occurs because our variability classification inherently involves some level of error when identifying binaries and RV standards, as we show in the top panel of Fig. \ref{fig:stsb}.

The identification of highly variable objects may also be influenced by the time interval between RV measurements. Therefore, we recognize that our thresholds for variability criteria could benefit from improvements. However, since the number of measurements per giant in our analysis is limited in the following section, imposing a stricter selection criterion on time between observations would reduce the total number of giants analyzed. Furthermore, variability is inherently constrained by the unknown set of orbital parameters that we do not attempt to resolve in this work. Nevertheless, using the two RVs with the lowest errors to define $w_{12}$ could help mitigate the issue of RV measurements being taken too closely in time.

\section{The RV variability of giants and their Li content}\label{sec:3}
We now turn to the central question of this work, which is to apply the methodology developed in Sect. \ref{sec:2} to determine whether there is any correlation between binarity and a high Li abundance among low-mass giants. As previously discussed, we compiled RV measurements from \textit{Gaia} DR3, GALAH DR3, and RAVE DR6. The \textit{Gaia} mission uses the RVS spectrograph with a resolving power of $R=\lambda/\Delta\lambda\sim11500$ for near-IR spectra. The \textit{Gaia} DR3 median formal precision for the RV data is on the order of $\sim1\ \mathrm{km\ s^{-1}}$ \citep[][]{Katz2023}. GALAH uses the HERMES spectrograph at the Australian Astronomical Observatory, reaching a resolution of $R\sim28000$ in the optical wavelength range. The RV uncertainties in the GALAH DR3 survey are typically $\lesssim0.1\ \mathrm{km\ s^{-1}}$ \citep[][]{Zwitter2021}. RAVE uses the 6dF multi-object spectrograph at the UK Schmidt telescope at Siding Spring in Australia, corresponding to a resolving power of $R\sim7500$ for a wavelength range of 8410–8795 Å. The RV uncertainties in RAVE DR6 are $<2\ \mathrm{km\ s^{-1}}$ \citep[][]{RAVEDR6}. \textit{Gaia} and GALAH report barycentric RV measurements, whereas RAVE provides heliocentric data. Since the difference between both frames is $\lesssim15\ \mathrm{m\ s^{-1}}$ \citep[e.g.,][]{Nidever2002, CollierCameron2021}, much less than the typical uncertainties of the RV measurements in the surveys, we ignore any correction related to the reference of the RV data.

The sample of giants to be analyzed is constructed as follows. We first took all stars in \textit{Gaia} DR3 with RV determinations, and crossmatched them with GALAH DR3 (so that Li measurements are available) and RAVE DR6, to produce a sample of stars with three independent RV determinations. Starting from this sample, we then selected giants following the recipe by \citet{Martell2021} (their Sect. 2.2), which uses the stellar parameters from GALAH. Then, we took those stars with no issues in their data, reduction, analysis, and iron abundance determination, summarized by the following flags in GALAH DR3: \texttt{flag\_sp == 0}, \texttt{flag\_fe\_h == 0}, extinction parameter E(B-V)<0.33 (to avoid difficulties with the analysis of spectra of stars with high extinction), and photometric quality flagged as A, B, or C in the WISE $W_{2}$ band (added in the catalog of GALAH) which is later used for classifying the evolutionary phase of the giants \citep[see][for further details of the parameters]{Buder2021}. Secondly, the sample is restricted to a range of effective temperature $3000$ K $\leq T_{\mathrm{eff}} \leq$ $5730$ K and surface gravity log(g)$\leq3.2$, which enclose the expected range for red giants in the first-ascendant RGB and the He-core burning phase. Next, we identified if there were any stars belonging to the SMC or LMC by using their spatial and kinematic properties as described by \citet{Martell2021}, but no members were found in our sample. 

We found a total of $5667$ red giants which have RV measurements with their respective errors reported by the three surveys. We also separately selected those stars that have \texttt{flag\_li\_fe == 0} in GALAH DR3, which indicates a line detection good enough to consider a Li abundance measurement without upper limits and flagged problems \citep{Martell2021, Buder2021}. This subset contains a total of $1418$ red giants. Li abundances were obtained using the logarithmic number density defined as A(X)=$\log(N_{X}/N_{H})+12$ for an element X. From the abundances available in GALAH, that is, [Li/Fe] and [Fe/H], we computed A(Li)=$[$Li/Fe$]$+$[$Fe/H$]$+1.05, where A(Li)$_{\odot}$=1.05 is the abundance value of the Sun from \citet{Asplund2009}. The errors for [Li/Fe] and [Fe/H] are given in GALAH DR3, then the errors for the Li abundance are obtained as eA(Li)=$\sqrt{\mathrm{e[Li/Fe]}^{2}+\mathrm{e[Fe/H]}^{2}}$. We found that the mean error is eA(Li)$\sim0.1$ and the maximum is eA(Li)$\sim0.25$. Since the A(Li) distributions are later analyzed with a bin size larger than the maximum error (see Sect. \ref{sec3.1}), we do not apply any additional selection based on eA(Li).

The range of metallicities in both samples is about $-1<$[Fe/H]$<0.5$, with an average of [Fe/H]$\sim-0.23\pm{0.24}$ for the sample of 5667 giants, and  [Fe/H]$\sim-0.26\pm{0.25}$ for the subsample of 1418 giants. Only $\lesssim4\%$ of the giants in the samples have metallicities  [Fe/H]$<-1$, where 2 of them are Li-rich. Although an anticorrelation between metallicity and fraction of close-binary systems has been observed in previous studies \citep[e.g.,][]{Moe2019}, we did not consider any specific selection or analysis related to different ranges of metallicity in our samples.

On the other hand, we also considered separating our sample with Li measurements by evolutionary phase. \citet{Martell2021}, based on the work of \citet{Sharma2018}, used a Bayesian stellar parameters estimator (\texttt{BSTEP}) to distinguish the stellar evolution phase of giants in GALAH DR3 between RGB and the red clump (RC). Table 1 of \citet{Martell2021} presents the evolutionary phase classification of $1262$ Li-rich giants including information on their positions (RA, DEC) and \textit{Gaia} DR2 identifier. We found $180$ coincidences between this sample and our giants with A(Li)>1.5. While they do not include the specific evolutionary phase per target in their public data for the non-Li-rich giants, the outputs of the \texttt{BSTEP} provide a probability of the giants to be in the RC, which is flagged in the public catalog of GALAH as \texttt{is\_redclump\_bstep}. Then, based on the argument that RC stars are standard candles and using the absolute magnitude $W_{2}$ of WISE, \citet{Martell2021} provides a method to identify the evolutionary phase: RGB when \texttt{is\_redclump\_bstep}$<0.5$ and $|W_{2}+1.63|>0.80$, and RC when \texttt{is\_redclump\_bstep}$\geq0.5$
and $|W_{2}+1.63|\leq0.80$. From this, we were able to separate by evolutionary phase 850 non-Li-rich giants.

\begin{figure}
    \centering
    \includegraphics[scale=0.6]{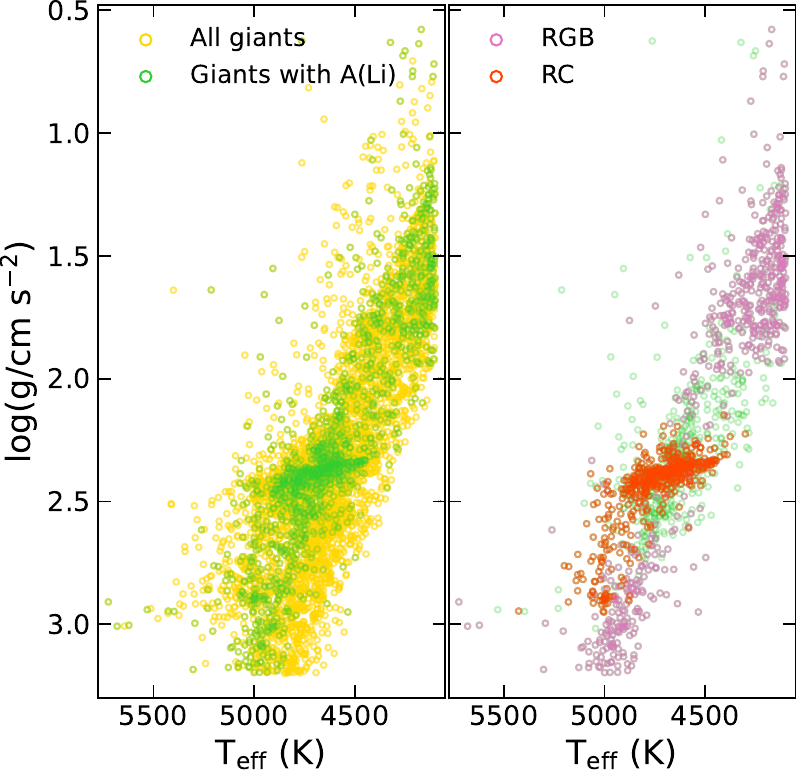}  \caption{The left panel shows the $\mathrm{log(g)}$-$\mathrm{T_{eff}}$ diagram for all the red giants found with RVs reported simultaneously in \textit{Gaia} DR3, GALAH DR3, and RAVE DR6 (yellow), and for those which also have a Li abundance reported in GALAH DR3 (green). The right panel shows the $\mathrm{log(g)}$-$\mathrm{T_{eff}}$ diagram for RGB (pink) and RC (orange) giants in our sample of giants with Li abundance (green background).}
    \label{fig:teff_logg}
\end{figure}

The left-hand panel of Fig. \ref{fig:teff_logg} shows that both samples, all the red giants and the subsample with Li abundances, are distributed in the range of $\log$(g) vs $\mathrm{T_{eff}}$ where the RGB and the RC are expected to have stars with a wide range of masses and metallicities. In the right-hand panel, we confirm that by showing the $1030$ giants we were able to separate by evolutionary phase (RGB or RC) along with the total sample with A(Li) measurements in the background.

At the same time, we expect that our sample must contain a fraction of giants in the asymptotic giant branch (AGB) since the \texttt{BSTEP} method does not distinguish this phase that can overlap in some cases with the RGB and RC in the log(g)-$\mathrm{T_{eff}}$ diagram. By considering that the evolutionary timescale of low-mass stars during the RGB plus the RC is about $\gtrsim500$ Myr, and in the AGB is $\sim5$ Myr \citep{LamersLevesque2017}, we estimate that the fraction of AGB stars in our sample is about $1\%$. 
Furthermore, some of these AGB giants could be Li-rich \citep[e.g.,][]{Reddy2005, Casey2016, Holanda2020}, \citet{DeepakReddy2019} reported that less than $5\%$ of their 335 Li-rich giants could be in the early-AGB, but these candidates can be also giants descending from the RGB tip to the RC. With this upper limit, we could expect a maximum of 9 of our Li-rich giants to be AGB.

Additionally, we use the WISE photometry \citep{Cutri2012} to test the  IR excess in our sample. For this, we consider the color $W_{1}-W_{4}>0.5$ as an indicator of IR excess. We found that $\sim30\%$ of the giants had IR excess regardless of their evolutionary stage or Li abundance. However, \citet{Rebull2015} discussed the implications of IR excesses and Li abundances using WISE data, pointing out that Li-rich giants with reported IR excesses may be artifacts in the catalog or sources of confusion. Then, by taking only the clean WISE data for the magnitudes $W_{1}$ and $W_{4}$ (they have the \texttt{cc\_flags} confusion flag set to \texttt{0000} and the \texttt{ph\_qual} photometric quality flag set to \texttt{A}), as done by \citet{Martell2021}, we found only 4 of 150 clean giants with IR excess, 2 of them in the RGB phase and 1 RC (the remaining target could not be classified by evolutionary stage). None of these clean giants are Li-rich.

\begin{figure*}
     \includegraphics[scale=0.565]{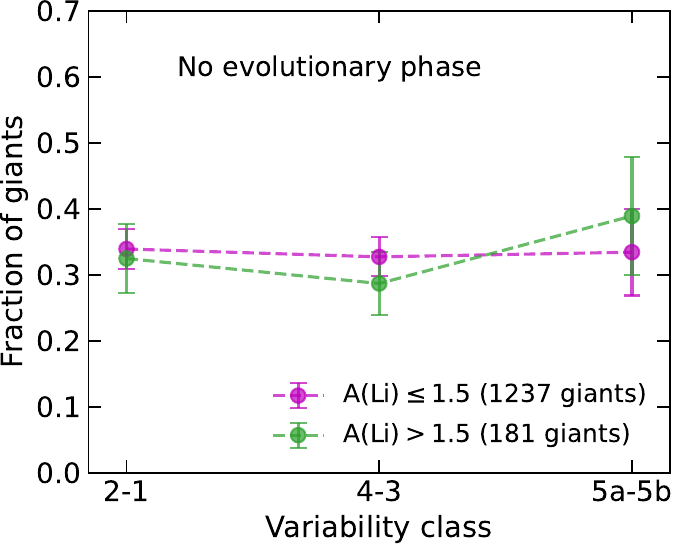}
     \includegraphics[scale=0.565]{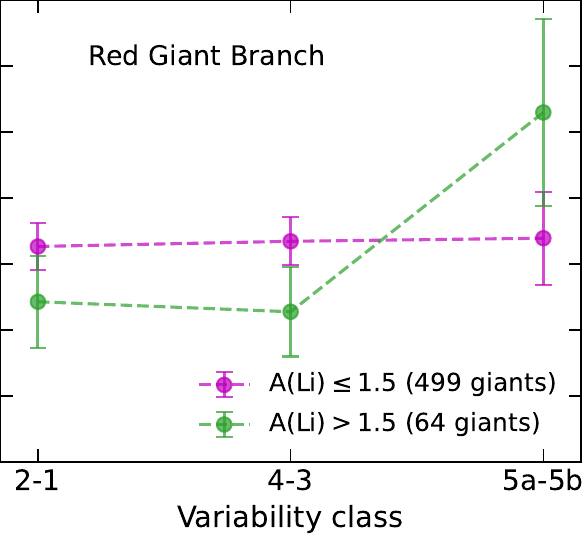}
     \includegraphics[scale=0.565]{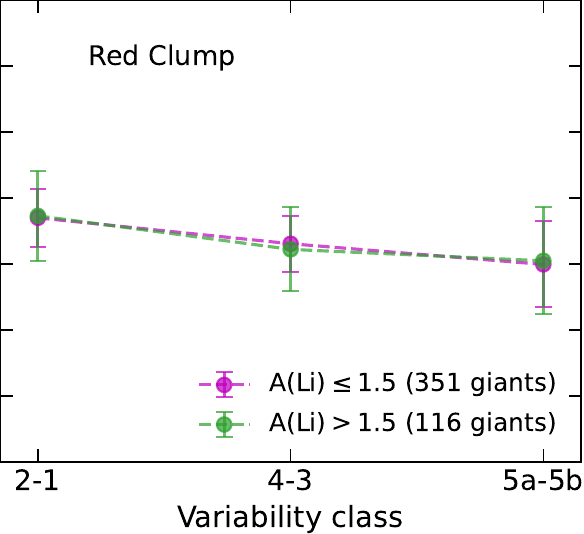}
    \caption{Fraction of RV-variable stars in our samples of Li-normal and Li-rich giants.  Insofar as the high-variability classes 5a-5b are indicative of binaries, the fraction of giants in the combined 5a-5b class can be thought of as the close binary fraction for the corresponding sample. In the left panel, no distinction is made between the evolutionary phases of the giants. The center panel shows only the first-ascent RGB giants, while the right panel displays only the RC giants. Evolutionary phases were determined following \citet{Martell2021}. There is a slight hint among first ascent (RGB) giants of higher RV variability (higher close binary fraction) for Li-rich than for Li-normal giants, although marginal given the estimated uncertainties,  When looking at the RC giants alone, however, the fraction of RV variables is the same among Li-rich and Li-normal stars.}
   \label{fig:lir_lin}
\end{figure*}

\subsection{RV variability and Li abundance}\label{sec3.1}
We applied our classification scheme of Sect. \ref{sec2.1} both to the 5667 giants with three RV epochs, and to the subsample of 1418 with good Li determinations from GALAH DR3. We used $RV_{1}=RV_{\mathrm{\textit{Gaia}}}$, $RV_{2}=RV_{\mathrm{GALAH}}$, and $RV_{3}=RV_{\mathrm{RAVE}}$. Table \ref{tab:tabvar} shows the fraction of giants in each variability class for both samples. We noted that the proportion of stars in each class is very similar for both subsets, that is, the giants with A(Li) measurement are a representative sample of the total set of giants with RV measurements registered in \textit{Gaia}, GALAH, and RAVE. In fact, even when the errors of the values in Table \ref{tab:tabvar} were taken to be a Poisson uncertainty of the counts for simplicity, all the fractions of the classes for the giants with A(Li) are within $<1\sigma$ of the fractions in the sample with all the giants.

An important caveat to consider is that some giants can exhibit pulsations, which can produce variations in RVs up to $\lesssim1\ \mathrm{km\ s^{-1}}$ and with typical values ranging from $\sim10$ to $100\ \mathrm{m\ s^{-1}}$ \citep[e.g.,][]{Hekker2008}. We note that $\Delta{RV_{\mathrm{max}}}=359\ \mathrm{m\ s^{-1}}$ is the minimum variation in our giants in classes 5a-5b, with only about $1.1\%$ and $8.4\%$ of them showing variations of $\Delta{RV_{\mathrm{max}}}<500\ \mathrm{m\ s^{-1}}$ and $\Delta{RV_{\mathrm{max}}}<1\ \mathrm{km\ s^{-1}}$, respectively. Thus, while we cannot conclusively distinguish between RV variations caused by pulsations and those induced by companion stars, only a small fraction are likely related to pulsations among our highly variable giants. A combination of both processes can also contribute to RV variations \citep{Beck2014}. Moreover, Cepheids and RRLyrae objects are among the most variable stars due to radial pulsations, with amplitudes $>10\ \mathrm{km\ s^{-1}}$. Given that \textit{Gaia} DR3 provides a catalog of such objects \citep{Ripepi2023}, we verified that our 1418 giants with measured Li abundances are not part of this catalog.   

\begin{table}
        \centering
        \caption{Percentage of giants in each variability class.}
        \label{tab:tabvar}
        \begin{tabular}{lccr} 
                \hline
        Class  & All giants (\%) & Giants with A(Li) (\%)\\
                \hline
        5a & $21.9\pm{0.6}$&$21.9\pm{1.2}$\\
        5b & $10.5\pm{0.4}$&$11.0\pm{0.9}$\\
        4  & $25.0\pm{0.7}$&$24.8\pm{1.3}$\\
        3  & $7.6\pm{0.4}$&$8.0\pm{0.8}$\\
        2  & $7.6\pm{0.4}$&$7.5\pm{0.7}$\\
        1  & $27.4\pm{0.7}$&$26.9\pm{1.4}$\\
                \hline
  
        \end{tabular}
\tablefoot{The sample includes all the giants with RV measurements in \textit{Gaia}, GALAH, and RAVE (5667 giants), and for the subsample with A(Li) measurements in GALAH (1418 giants).}
\end{table}

Now, we combine the A(Li) information with the variability classes obtained to explore if there is any relation between Li content and RV variability.  
We illustrate in Fig. \ref{fig:lir_lin} the fractions of giants as a function of our variability classes, for Li-normal (A(Li)$\leq$1.5) and Li-rich giants separately\footnote{We included the level of uncertainty of our variability classification by considering the results of the tests made in Sect. \ref{sec2.2}. Since we found about $p_{nv}=0.802$ and $p_{v}=0.849$ of effectiveness for classifying non-variables and variables respectively, we calculate the expected value of correctly detected non-variables to be $p_{nv}n_{nv}$ for $n_{nv}$ giants found in classes $\leq4$. Consequently, $p_{v}n_{v}$ is the value of correctly detected variables for $n_{v}$ giants found in classes 5a-5b. These values of effectiveness have an intrinsic uncertainty associated. We add that contribution to the error of the expected values for non-variables or variables by taking $\sigma_{g}=\sqrt{\sigma_{e}^{2}+\sigma_{p}^{2}}$, where $\sigma_{g}$ is the error on the expected number of giants in a certain variability type, $\sigma_{e}=\sqrt{pn}$ the Poisson error for $n$ measurements with individual probabilities $p$ to obtain the expected value, and $\sigma_{p}=ne_{p}$ for $e_{p}$ the error of $p$. From the tests presented in Sect. \ref{sec2.2} we estimated $e_{p,nv}=0.05$ for the non-variables and $e_{p,v}=0.16$ for the variables.}. In this case, to avoid clutter, we consider joint classes, that is, class 5a-5b, class 4-3, and class 2-1 (for simplicity we usually refer in this analysis as non-variables the giants in classes $\leq4$, and variables to those giants with high certainty of variability, i.e., classes 5a-5b).  In the left-hand panel,  we present the expected fraction of giants in each variability class for the Li-normal and Li-rich giants without considering the evolutionary phase. It can be seen that the fraction of Li-rich giants is essentially the same as that of Li-normal giants for all RV variability classes, given the error bars.

In the center panel of Fig. \ref{fig:lir_lin} is the expected fraction of RGB stars in each variability class. In this case, the Li-rich RGB population has a larger fraction of stars in the high variability class than the $\sim33\%$ found in the Li-normal RGB and the samples without considering the evolutionary phase. Consequently, the fraction of non-variable Li-rich giants is smaller than those in the other Li-normal samples. However, the total number of Li-rich RGBs is the smallest sample analyzed in Fig. \ref{fig:lir_lin}, and as such the errors of the fractions are also large. This means that the error bars of the fractions in each variability class are blended at least within one standard deviation with the non-Li-rich RGBs. This apparent difference, however, goes away completely among the subsample of RC giants (right-hand panel), and we note that not only is the fraction of Li-rich RC giants the same as that of Li-normal RC giants for all RV-variability classes, but the fraction of Li-rich giants with large RV variability has decreased from 52\% in the RGB to 30\% in the RC. Even though our estimated error bars may account for this difference between Li-rich giants in the RGB and the RC, we think it is something worth following up with further work.

\begin{figure*}
    \centering
    \includegraphics[scale=0.64]{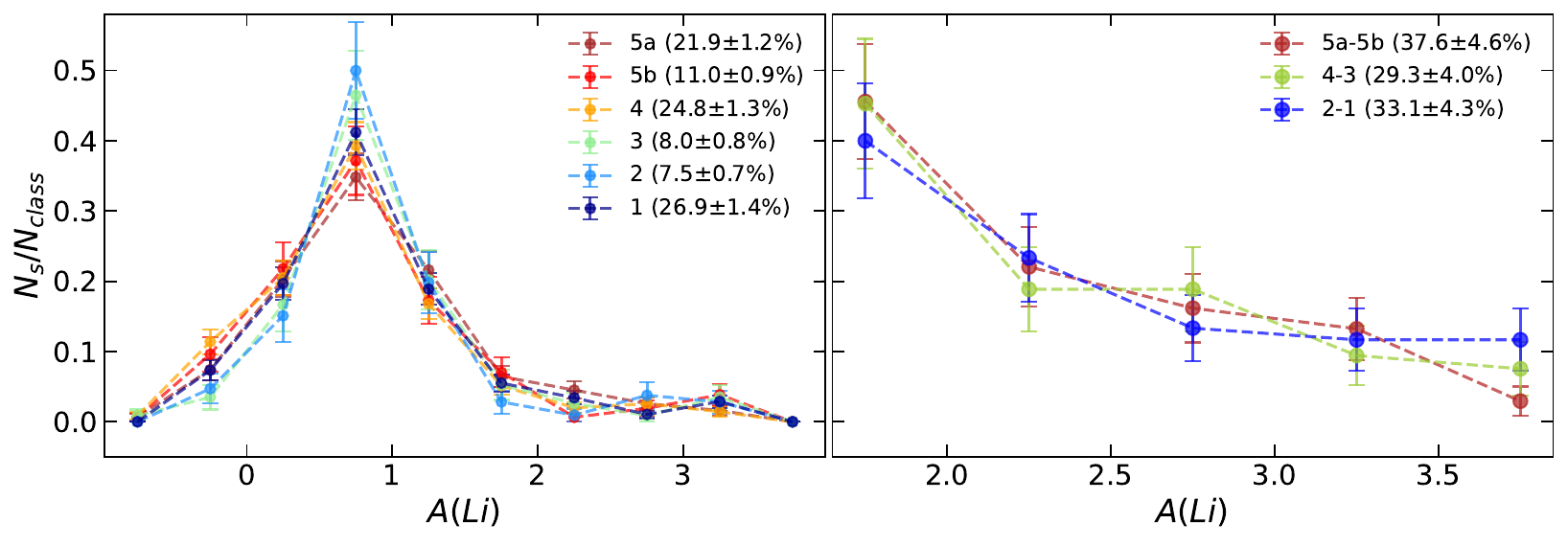}
    \caption{The left panel shows the A(Li) distribution for each class of variability obtained for the giants. We note that $\mathrm{N_{s}/N_{class}}$ is the fraction of stars in a class for the corresponding bin of A(Li). The percentages represent the fraction in each class just like in Table \ref{tab:tabvar}. The right panel shows the A(Li) distribution for each joint class of variability obtained for the Li-rich (A(Li)>1.5) giants. For this case, the variability is high for class 5a-5b, medium for class 4-3, and low for class 2-1; the percentages represent the fraction of giants in each joint class for the total Li-rich giants with errors obtained following the same method applied in Table \ref{tab:tabvar}.}
    \label{fig:li_dist}
\end{figure*}

To get greater insight into the relationship between A(Li) in giants and variability we analyze the A(Li) distribution for all the variability classes defined under the $w$-criteria. The left panel of Fig. \ref{fig:li_dist} shows the A(Li) distribution of each class obtained from the $w_{12}$-$w_{t}$ criterion for our sample of giants. The vertical axis shows the number of stars $\mathrm{N_{s}}$ in a given A(Li) bin of range 0.5 dex (starting from A(Li)=$-1$), normalized by $\mathrm{N_{class}}$, the total number of stars with Li determination that are in the corresponding variability class. The first immediate conclusion from Fig. \ref{fig:li_dist} is that all variability classes follow almost the same distribution, which is very similar to a normal distribution. Moreover, they all reach their maximum value at the same bin of Li abundance: A(Li) centered at 0.75 dex. This means that there are no obvious differences regarding Li content among all the variability classes, which we are using as a proxy for the likelihood of binarity. There is a hint that the peak of the A(Li) distribution may be slightly wider for the most variable classes (4, 5a, and 5b). But, given the size of the error bars, these differences are not larger than $1\sigma$, and therefore not significant.

The right-hand panel of Fig. \ref{fig:li_dist} zooms into the high-end tail of the A(Li) distributions from the left-hand panel, starting from A(Li)=1.5. In this panel, we consider joint classes. Thus, $\mathrm{N_{s}/N_{class}}$ is the fraction of stars in a joint class for a given range of A(Li). Here we consider the fractions in the bins normalized with respect to the Li-rich giants in each class. Consequently, given the different normalizations, the shape of the distributions exhibits slight variations compared to what is observed in the left-hand panel of the figure. Giants in class 5a-5b are therefore those Li-rich giants with high certainty of being in a binary system (because they have the largest variation in RV), and they correspond to $37.6\%$ of all the Li-rich stars in our sample. This is only about $5\%$ larger than the fraction of stars with high certainty of variability when considering all giants with A(Li) measurement ($32.9\%$) and all giants with RV in the three surveys ($32.3\%$). Considering error bars, $33\%$ is about one standard deviation from the fraction of Li-rich giants in class 5a-5b (37.6\%) which is marginal. Therefore, regardless of their Li abundance, there are about one-third of giants with a high probability of being in a binary system. At the same time, among the Li-rich giants only, there are no significant differences among the A(Li) distributions, they all show that the fraction $\mathrm{N_{s}/N_{class}}$ of giants in each A(Li) bin is almost the same for all the variability classes.

These results are consistent with previous work with smaller samples of red giants but more RV measurements per star. \citet{Jorissen2020} followed up a set of 24 red giants for about 9 years and less than $36\%$ of the Li-rich giants were identified with a high probability of being in binary systems under the F2-distribution method. \citet{AguileraGomez2023} also compared the A(Li) distribution of a handful of field red giants, of which some were confirmed as spectroscopic binaries, and others with a planet, and they also did not find any trend between a companion and the Li enrichment. Other correlations have been studied that could hint toward binary interaction producing very Li-rich giants, for example, with $[$C/N$]$ ratio \citep{Tayar2023}. This would be an additional method that could be explored to understand if binary interactions are responsible for all or a subset of Li-rich giants. Additionally, extensive follow-up of Li-rich giants and measurements of multiple RV epochs is necessary to confirm the results presented in this work.

We also tested our criteria for RV variability using the 24 giants from \citet{Jorissen2020}. In their work, each star has multiple RV measurements. For each giant, we designated the measurement with the largest RV error as \( RV_{3} \) and explored all possible combinations with the remaining RVs to determine \( RV_{1} \) and \( RV_{2} \). Among the 11 definitive and possible binaries identified in their sample, 9 were consistently classified as class 5a-5b in more than 50\% of their RV combinations, with 7 classified as such in over 90\% of combinations. For the 13 non-variables, all were classified in classes $\leq$4 in more than 50\% of their RV combinations, with 8 classified as such in over 90\% of combinations. This result indicates that our variability criteria can reproduce similar results to previous studies, although limited by the number of measurements used for our classifications.

\subsection{Super Li-rich giants.}

While A(Li)>1.5 is the typical threshold for considering Li-rich giants, there is a fraction of objects that have Li abundances higher than the expected value from the Big Bang nucleosynthesis, that is, A(Li)>2.7 \citep{Cyburt2008, Fields2020}, and some with higher values than the Solar System meteoritic value \citep{Lodders1998}, that is, A(Li)>3.3. These are often referred to as super Li-rich giants. Such abundances could require internal Li production to be reached (or probably a combination of internal and external processes). The interaction with a binary companion has been proposed as one of the processes that can favor the Li production and transport to the surface in giants \citep{Casey2019}, then it is worth studying the fraction of super Li-rich giants in our sample.

\begin{table}
        \centering
        \caption{Giants in class 5a-5b and class $\leq$ 4 for different Li-rich ranges.}
        \label{tab:super_li}
        \begin{tabular}{lccr} 
         \hline
        Total  & A(Li)>1.5 & A(Li)>2.7 & A(Li)>3.3\\
                \hline
                Class 5a-5b & 68 (37.6\%) & 17 (32.1\%) & 4 (22.2\%)\\
                Class $\leq$4 & 113 (62.4\%) & 36 (67.9\%) & 14 (77.8\%)\\\\
                
                \hline
       RGB  & A(Li)>1.5 & A(Li)>2.7 & A(Li)>3.3\\
                \hline
                Class 5a-5b & 33 (51.6\%) & 9 (50.0\%) & 3 (60.0\%)\\
                Class $\leq$4 & 31 (48.4\%) & 9 (50.0\%) & 2 (40.0\%)\\\\
                \hline
         RC  & A(Li)>1.5 & A(Li)>2.7 & A(Li)>3.3\\
                \hline
                Class 5a-5b & 34 (29.3\%) & 8 (22.9\%) & 1 (7.7\%)\\
                Class $\leq$4 & 82 (70.7\%) & 27 (77.1\%) & 12 (92.3\%)\\
                \hline
  
        \end{tabular}
 \tablefoot{The number of giants in each class is presented, while the percentages in parentheses indicate the fraction of giants in each class relative to the total number of giants in the range of A(Li). The table is divided into three sections to identify the number of Li-rich giants in each variability class during specific evolutionary phases (RGB or RC), based on the cross-match with the classified GALAH DR3 giants from \citet{Martell2021}.}
\end{table}
Table \ref{tab:super_li} shows the amount of Li-rich and super Li-rich giants that are identified as class 5a-b and class $\leq4$ for the total sample and those classified as RGB and RC separately. We found that the fraction of giants with high variability among super Li-rich giants with A(Li)>2.7 is also about one-third, and it is <$25\%$ for those with A(Li)>3.3 without taking into account the evolutionary phase, which is consistent with what we discussed in Sect. \ref{sec3.1} where independent of the range of Li considered the fraction of giants with high variability is in a tight range.

\begin{figure*}   
    \centering
    \includegraphics[scale=0.67]{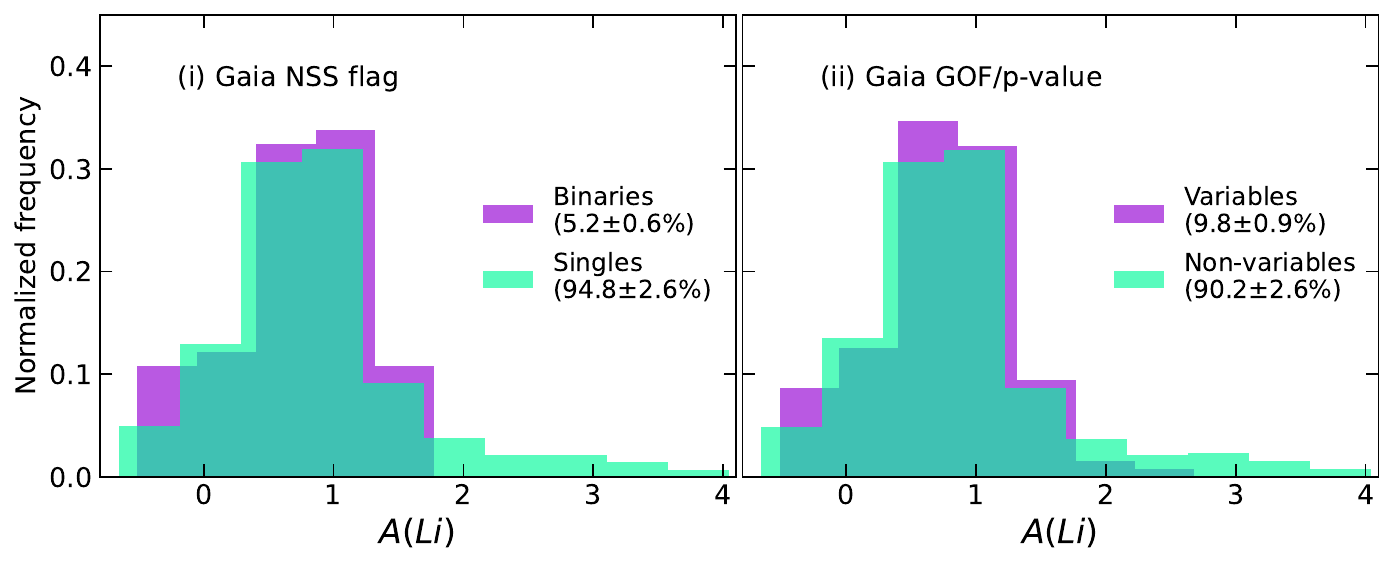} 
  \caption{First panel (i): A(Li) distribution for the red giants identified as binaries and singles using the non-single star (NSS) flag provided by \textit{Gaia} DR3. Second panel (ii): A(Li) distribution for the red giants identified as variables and non-variables using the p-value for constancy and the radial velocity renormalized goodness of fit (GoF) reported in \textit{Gaia} DR3 following the method described by \citet{Katz2023}.}
    \label{fig:gaia_bin}
\end{figure*}

When considering the evolutionary phase, for all the Li abundance ranges shown in Table \ref{tab:super_li}: about $35\%$ of the Li-rich giants are RGB, while about $65\%$ are RC, which is consistent with what was found by \citet{Martell2021}. However, regardless of the Li abundance range considered, about $50\%$ of the (super) Li-rich RGB stars show high certainty of variability. In contrast, the Li-rich RC stars show lower proportions of high variability than the total sample for the three A(Li) ranges shown. 

These results require further investigation since the total number of super Li-rich giants is small (especially for the RGB case). It is important to note, for example, that if we consider Poisson deviations in the data of Table \ref{tab:super_li}, the errors of the fractions of super Li-rich RGB (A(Li)>2.7 and >3.3) are about $15\%$-$30\%$. Further, we previously estimated that about 9 of the Li-rich giants in our sample could be in the AGB phase. Some of them are likely to be super Li-rich since depending on the mass, the extension of the convective envelope during the AGB pulsating phase can rapidly transport the Li to the surface, allowing it to reach values A(Li)$>2.7$ \citep[see][]{VenturaDantona2010, Lau2012}.

\section{Binarity and variability in \textit{Gaia} DR3}\label{sec:4}

In this section we explore the feasibility of our methods for detecting RV variability and the results when analyzing the A(Li) distribution of different classes using indicators of binarity and RV variation reported in the last data release of \textit{Gaia}.

\textit{Gaia} DR3 offers an identification for non-single systems primarily based on orbital resolution \citep{Hambly2022}. A \texttt{non\_single\_star>0} (NSS>0) flag in \textit{Gaia} indicates that the methods have detected a binary system. We tested this flag in our set of 1418 and obtained that $5.2\%$ of them have NSS>0, significantly different from the $32.9\%$ with high variability based on our $w$ values. \citet{Gaia2023} and \citet{Pourbaix2022} have pointed out that the NSS catalog of \textit{Gaia} is still incomplete, so we need to keep this caveat into account. We estimate a degree of incompleteness by testing the NSS flag on our SB9 stars and found that only $60.6\%$ of them have NSS>0. Then, based on the tests made in Sect. \ref{sec:2}, our $w$ parameters are likely classifying more close binaries than the NSS flag in \textit{Gaia} DR3. In Panel (i) of Fig. \ref{fig:gaia_bin} we also show the Li abundance distributions for single stars and binary systems found with the NSS flag. The two distributions are alike, except that almost all the Li-rich giants are not binaries and there are no NSS>0 systems with A(Li)>1.8. Even considering the incompleteness of the NSS flag, these results suggest an agreement with what we obtained using the $w$ parameters, that is, Li-rich giants should not be mostly related to binary systems.

An alternative way of detecting RV variability from parameters in the \textit{Gaia} DR3 catalog was reported by \cite{Katz2023}: a star with normalized goodness of fit (GoF) \texttt{rv\_renormalised\_gof > 4}, and consistency for RV measurements \texttt{rv\_chisq\_pvalue $\leq$ 0.01} (p$\leq$0.01), is variable\footnote{The GoF describes an empirical quantity that represents the scatter of RVs when compared to the typical epoch uncertainty. We note that p is the probability that the $\chi^{2}$ for the RVs exceeds an expected value of the probability distribution with a certain significance \citep[see][]{Hambly2022}.}. When testing these criteria in our SB9 stars, $92.0\%$ of them are identified as variables.  This is a similar performance as the $w$-criteria for detecting binaries ($\sim85$-$94\%$). We use this alternate method for our giants and compare the A(Li) distributions obtained for variables and non-variables in the second panel (ii) of Fig. \ref{fig:gaia_bin}. These distributions appear to be quite similar to what is shown in panel (i). Furthermore, just $9.8\%$ of the giants were identified as variables, and the maximum Li abundance identified for this subsample is A(Li)$\sim$ 2.7. This disagrees with the about $33\%$ of variable giants found by our method and also by \citet[][]{Badenes2018} when analyzing APOGEE giants.

The discrepancies between our w-criteria and the method of \citet{Katz2023} are likely due to the tight criteria for variability the latter considers: it requires analyzing more than ten epochs. Thus, to relax these criteria to something more comparable to our $w$ method, we obtained the values of p and GoF using the 3 epochs of RV used in Sect. \ref{sec:3} and following the description of \citet{Hambly2022}. In doing so, similarly to the previous tests, 10.4\% of the giants were identified as variables. Furthermore, the formulation of the GoF is almost the same as the F2 distribution \citep[e.g., see][]{Jofre2023}\footnote{Except that we should include a normalization factor mode(UWE)$\sim$ 1 for a range $\mathrm{T_{eff}} \approx 4000$-$5800$ K \citep[see][]{Sartoretti2022}.}. Then, by relaxing p in the description of \citet{Katz2023} we should obtain a proxy of the F2 distribution. We noted that as we relax p, we classify more giants as variables, and when using just the GoF, we obtain 38.2\% of the giants being variables and with an A(Li) distribution almost identical to what we obtained when using the w-criteria.

\section{Summary and conclusions}\label{sec:5}

We developed a method for detecting RV variability when using RVs from heterogeneous sources, to apply it to low-mass field giants with a well-determined Li content and explored the question of whether the phenomenon of Li-rich giants is a result of binary stellar evolution. The method is a variation of the F2 statistic and works by comparing three independent RVs from different epochs, accounting for the corresponding uncertainties. We tested and calibrated the sensitivity of the method against two control samples, a catalog of RV standard stars and another of confirmed spectroscopic binaries, to evaluate the degree of confidence in identifying single (RV non-variable) and binary (RV variable) stars. We find, non-surprisingly, that the accuracy of the method in correctly separating the two control samples mainly depends on the precision of the RVs available for the single stars.  The separation achieves a 100\% success rate when only high-precision RVs are used for the RV standards (the "single" stars), and diminishes to 80-85\% when the less precise RVs from RAVE and \textit{Gaia} are included in the RV variability analysis (see Figs. \ref{fig:hist} and \ref{fig:stsb}). Since these are the data that we will use for the analysis of giants with a Li content, this is the level of certainty that our ability to identify binaries among our sample of giants will have. We also compared our results with the information on binarity and variability available in \textit{Gaia} DR3, but find that the available information is insufficient for our purposes because the survey has thus far only flagged a small fraction of binary systems.

We assembled all stars with RV determinations from \textit{Gaia} DR3, RAVE DR6, and GALAH DR3, and, following the recipe from the GALAH team itself, we used stellar parameters from the latter to make a selection of 1418 giants with good Li determinations. For 1030 of these, we could determine evolutionary states, producing subsamples of 563 first ascent red giants (RGBs) and 467 red clump giants (see Fig. \ref{fig:teff_logg}). This is the largest sample of low-mass giants with a well-determined Li content ever studied for binarity purposes. We applied our RV variability classification method to these samples, and our main findings are summarized hereafter.

\begin{enumerate} 
\item The fraction of giants (RGB+RC) with a high certainty of RV variability is $\sim$33\%, which is on the order of the close binary fraction observed for main-sequence stars \citep[e.g.,][]{Gao2017, Moe2019} and giants \citep[][]{Badenes2018} at typical disk metallicities. This provides evidence, in addition to that obtained from the control samples, that our classification analysis is reliable.

\item The fraction of RV variable stars in the full sample (RGB+RC; left-most panel in Fig. \ref{fig:lir_lin}) is about the same for Li-rich giants (A(Li) $>$ 1.5) and for Li-normal giants (A(Li) $<$ 1.5).  Moreover, there is no excess of RV variable Li-rich giants with an extreme Li content (the super Li-rich giants, with A(Li) $>$ 2.7 or A(Li) $>$ 3.3).  All this argues against a binary interaction phenomenon as the main mechanism for explaining Li-rich giants in general.

\item The fraction of Li-rich giants is more than two times larger in the RC than in the RGB.  Although we did not attempt to quantify selection effects in our samples with respect to the Li content, we note that this is consistent with the fact that most Li-rich giants with seismic-based determinations of mass and evolutionary state have turned out to be clump giants \citep[e.g.,][]{SilvaAguirre2014, DeepakReddy2019, DeepakLambert2021, Mallick2023}. Even so, the incidence of Li-rich objects among first-ascent RGB giants in our sample ($\sim$ 11\%) is not small, and they too remain unexplained by standard stellar evolution.

\item When focusing on the RC population (right-most panel in Fig. \ref{fig:lir_lin}), we find that the fractions of Li-rich and of Li-normal RC giants are the same across all RV variability classes, again likely indicating that the close binary fraction of Li-rich giants in the RC is not special.  This is a highly significant result statistically speaking, obtained for the largest sample of RC giants with a well-determined Li content ever analyzed for these purposes, and this keeps adding to the evidence against a binary-related genesis for Li-rich giants in the RC.  For super Li-rich RC giants, the statistics become poorer, but we observe that the fraction of highly RV-variable objects of that type appears to decrease rapidly with an increasing Li content.

\item While the binary fraction in the RC does not appear to depend on the Li content, we find evidence that this may not be the case for first-ascent RGB giants, for which we determine that about half of the Li-rich and super Li-rich objects in this evolutionary state have a high RV variability (middle panel in Fig. \ref{fig:lir_lin}).  This is an intriguing result that contrasts with what we find for the RC, and which may be signaling some causal connection between binarity and Li production during the first-ascent RGB.  However, as there are only half as many RGB giants as RC giants in our sample, this is a result not as statistically significant as that of the RC.  Future work with a larger number of RV epochs and a cadence better suited for binary detection and characterization would be needed to confirm or rule out this specific result with a higher degree of confidence.

\end{enumerate} 

Thus the main conclusion of the present work needs to be separated by the evolutionary state.  In the case of Li-rich giants in the clump, our RV variability analysis argues against binaries having to do with the origin of the excess Li.  This lends support to work that provides indirect but compelling evidence for a causal relation with the core helium flash (\citealt{Mallick2023}; del Moral \& Chanamé, in prep.).  In the case of Li-rich giants on the first-ascent RGB, instead, our analysis leaves some open room for a binary-related origin of the excess Li, but more RV measurements are needed for more robust conclusions for that evolutionary state.  We stress that this apparent dichotomy between Li-rich objects in the RC and the RGB should not necessarily be seen as a problem, because, as discussed in \citet{Chaname2022}, today's field red giants are different than the progenitors of today's field red clump giants.  Field clump giants are a younger population, with a stellar mass distribution that extends to higher masses than the older field red giants we observe today, and thus it may well be the case that Li-rich giants on the RGB have a different origin than Li-rich giants in the clump.

\begin{acknowledgements}
M.C.-T. acknowledges support from the Trottier Space Institute (TSI) and the Centre de Recherche en Astrophysique du
Québec (CRAQ). J.C. acknowledges support from the Agencia Nacional de Investigaci\'on y Desarrollo (ANID) via Proyecto Fondecyt Regular 1231345; and from ANID BASAL projects CATA-Puente ACE210002 and CATA2-FB210003. 
We thank the anonymous referee for the insightful suggestions on the previous versions of the paper.
\end{acknowledgements}
\bibliographystyle{aa}

\end{document}